\documentclass[superscriptaddress,amsmath, amssymb, amsfonts, twocolumn, footinbib]{revtex4-2}
\usepackage[german,american,english]{babel}
\usepackage{graphicx}
\usepackage{graphics}
\usepackage{dcolumn}
\usepackage{multirow}
\usepackage{soul}
\usepackage{bm}
\usepackage{latexsym}
\usepackage{amssymb}
\usepackage{amsmath}
\usepackage{amsfonts}
\usepackage{layout}
\usepackage{verbatim}
\usepackage{epsfig}
\usepackage{graphicx}
\usepackage{amsbsy}
\usepackage{lipsum}
\usepackage{cancel}
\usepackage[toc]{appendix}
\usepackage{xcolor}

\newcommand{\bea}{\begin{eqnarray*}}
	\newcommand{\eea}{\end{eqnarray*}}
\newcommand{\bne}{\begin{equation*}}
\newcommand{\ede}{\end{equation*}}
\newcommand{\ba}{\arraycolsep 0.3ex \begin{array}{rl}}
\newcommand{\ea}{\end{array}}

\newcommand{\bnen}{\begin{equation}}
\newcommand{\eden}{\end{equation}}
\newcommand{\bean}{\begin{eqnarray}}
\newcommand{\eean}{\end{eqnarray}}
\newcommand{\bsen}{\begin{subequations}}
	\newcommand{\esen}{\end{subequations}}

\newcommand{\bna}{\begin{array}}
	\newcommand{\eda}{\end{array}}
\newcommand{\bnm}{\begin{enumerate}}
	\newcommand{\edm}{\end{enumerate}}

\def\pz{{\partial}}

\def\HBd{\hat{\bm d}}

\def\Hj{{\hat \jmath}}

\def\Hv{{\hat v}}

\def\Hr{{\hat r}}
\def\HBr{\hat{\bm r}}

\def\HL{{\hat L}}

\def\Bk{{\bm k}}
\def\Bh{{\bm h}}
\def\BE{{\bm E}}
\def\HH{{\hat H}}
\def\CR{{\mathcal R}}
\def\BCR{{\bm\CR}}
\def\Hrho{{\hat\rho}}

\def\RD{{\rm D}}
\def\RF{{\rm F}}
\def\RR{{\rm R}}
\def\AFM{{\rm AFM}}

\def\eps{\epsilon}
\def\ve{{\varepsilon}}
\def\Bsigma{{\bm\sigma}}

\def\frac#1#2{{\textstyle{#1 \over #2}}}
\def\der#1#2{{\pz #1\over\pz #2}}
\def\Der#1#2{{D #1\over D #2}}

\def\Tra{\mathop{\textsf{Tr}}}
\def\nd{^{\vphantom{\dagger}}}

\def\half{\frac{1}{2}}

\def\hh{\hskip 0.1em}

\def\sbraket#1#2{{\langle \hh #1  \hh |  \hh #2 \hh  \rangle}}

\def\ket#1{{\big| \hh #1\hh \big\rangle}}

\usepackage[utf8]{inputenc}

\begin{document}
	
\title{Quantum corrections to the orbital Hall effect}

\author{Hong Liu}
\altaffiliation{These authors contributed equally to this work.}
\affiliation{School of Physics, The University of New South Wales, Sydney 2052, Australia}
\affiliation{ARC Centre of Excellence in Low-Energy Electronics Technologies, UNSW Node, The University of New South Wales, Sydney 2052, Australia}
\author{James H. Cullen}
\altaffiliation{These authors contributed equally to this work.}
\affiliation{School of Physics, The University of New South Wales, Sydney 2052, Australia}
\author{Daniel P. Arovas}
\affiliation{Department of Physics, University of California at San Diego, La Jolla, California 92093, USA}
\author{Dimitrie Culcer}
\affiliation{School of Physics, The University of New South Wales, Sydney 2052, Australia}
\affiliation{ARC Centre of Excellence in Low-Energy Electronics Technologies, UNSW Node, The University of New South Wales, Sydney 2052, Australia}

\begin{abstract}
Evaluations of the orbital Hall effect (OHE) have only retained inter-band matrix elements of the position operator. Here we evaluate the OHE including all matrix elements of the position operator, including the technically challenging intra-band elements. We recover previous results and find quantum corrections due to the non-commutativity of the position and velocity operators and inter-band matrix elements of the orbital angular momentum. The quantum corrections dominate the OHE responses of the topological antiferromagnet CuMnAs and of massive Dirac fermions.
\end{abstract}

\date{\today}
\maketitle

{\color{blue}\textit{Introduction.}} The non-equilibrium properties of Bloch electrons' orbital angular momentum (OAM) \cite{Yafet-1963,Vanderbilt_2018} have come under intense scrutiny with the advent of orbitronics \cite{Orbitronics-PRL-2005-Shoucheng, Orbitronics-in-action, Rhonald-Rev}, whose focus is generating non-equilibrium OAM densities and currents \cite{Exp-OHE-Ti-Nat-2023-Hyun-Woo,Hong-OHE-PRL,OHE-Binghai,OHE-PRB-2022-Manchon,OHE-metal-PRM-2022-Oppeneer,OHE-BiTMD-PRL-2021-Tatiana,RS-OHE-disorder,ISOHE-PRL-2018-Hyun-Woo,IOHE-Metal-PRB-2018-Hyun-Woo,OHE-Hetero-PRR-2022-Pietro,OHE-Weak-SOC-npj, PhysRevLett.131.156702, PhysRevB.106.184406, PhysRevLett.131.156703,10.1063/5.0106988,Exp-OEE-PRL-2022-Jinbo, voss2024nonequilibriumorbitaledgemagnetization}. The orbital Hall effect (OHE) refers to the generation of a transverse OAM current by an electric field. The technological motivation underpinning OHE efforts is the electrical manipulation of magnetic degrees of freedom \cite{OT-FM-PRB-2021-YoshiChika,OT-OEE-NatComm-2018-Haibo,OT-PRR-2020-Hyun-Woo,OT-NatComm-2021-Kyung-Jin, Exp-OT-PRR-2020,Exp-OT-CommP-2021-Byong-Guk, LS-conversion-CP-2021-Byong-Guk,OOS-Cvert-2020-PRL-Mathias, PhysRevResearch.4.033037, OHE-OT-large, L-S-OT-2023}, with an emphasis on weakly spin-orbit coupled materials \cite{Exp-graphene-OHE-arXiv-2022, OAM-Exp-Tobias,OAM-PRL-2020-Reinert, OEE-NatComm-2019-Peter, Tangping, Inverse-OHE-weak-SOC,PhysRevResearch.6.013208}. From a fundamental science perspective, the OHE has also been proposed as one of the mechanisms behind the observed valley Hall effect \cite{IOHE-PRB-2021-Giovanni,OAM-VHE-MoS2-NatPhy-2013-Xiaodong,OAM-VHE-SciRep-2015-Jinbo,OAM-VHE-MoS2-NatPhy-2013-Xiaodong,TMDsoc-IOHE-PRB-Satpathy-2020,VOHE-TMD-PRB-Paul-2020}, and attempts to disentangle these two effects are now underway \cite{OHE-BiTMD-PRL-2021-Tatiana, BiTMD-OHE-PRB-2022-Giovanni&Tatiana,PhysRevLett.132.106301-Giovanni}. 

The OAM and OHE involve the position operator, which is challenging in extended systems \cite{PhysRevB.108.245105,OEE-scalar-potential,OEE-NL-2018-Shuichi,OEE-SciR-2015-Shuichi,PhysRevLett.108.046805,OAM-PRB-2013-JungHoon,OAM-Rashba-PRL-2011-Changyoung, Resta-PhysRevLett.95.137205, Resta-PhysRevB.74.024408, Resta-PhysRevResearch.2.023139, Position-operator-Rev}. In particular, the band-diagonal elements of the position operator are themselves differential operators, which act on the quantities surrounding them. This difficulty can be circumvented in equilibrium, since the equilibrium OAM expectation value involves only inter-band matrix elements of the position operator. In light of this, the conventional evaluation of the OHE, which includes earlier work by some of us \cite{Hong-OHE-PRL}, has proceeded as follows: (i) start with the OAM matrix elements appearing in the equilibrium OAM expectation value; (ii) multiply these by the velocity operator matrix elements; and (iii) combine this product with a non-equilibrium distribution found using standard methods such as the Boltzmann equation or the Kubo formula \cite{Rhonald-Rev,Hong-OHE-PRL,BiTMD-OHE-PRB-2022-Giovanni&Tatiana, IOHE-PRB-2021-Giovanni,OHE-BiTMD-PRL-2021-Tatiana,OHE-PRB-2022-Manchon,PhysRevB.108.075427-Manchon,OM-Berry-PRB-2016-Mokrousov,CIAM-PRR-2020-Yuriy,ISOHE-PRL-2018-Hyun-Woo,IOHE-Metal-PRB-2018-Hyun-Woo,IOHE-XIV-PRB-2021-Hyun-Woo,OT-FM-PRB-2021-YoshiChika,Tangping,Inoue-SHE-PhysRevB.70.041303,Inoue-AHE-PhysRevLett.97.046604,doi:10.1021/acs.nanolett.3c05129, PhysRevLett.130.116204, OHE-Topology-2023}. We argue that this approach is incomplete and misses important terms, a fact that can be seen already at the operator level. Replacing the OAM operator appearing inside the orbital current by the equilibrium OAM matrix elements is inappropriate, as it amounts to neglecting the band-diagonal matrix elements of the position operator. These matrix elements, being differential operators, act on the velocity matrix elements surrounding them. The additional resulting terms are substantial, and in certain cases overwhelm the conventional terms in the OHE.   

\begin{figure}[t]
\begin{center}
\includegraphics[trim=0cm 0cm 0cm 0cm, clip, width=0.9\columnwidth]{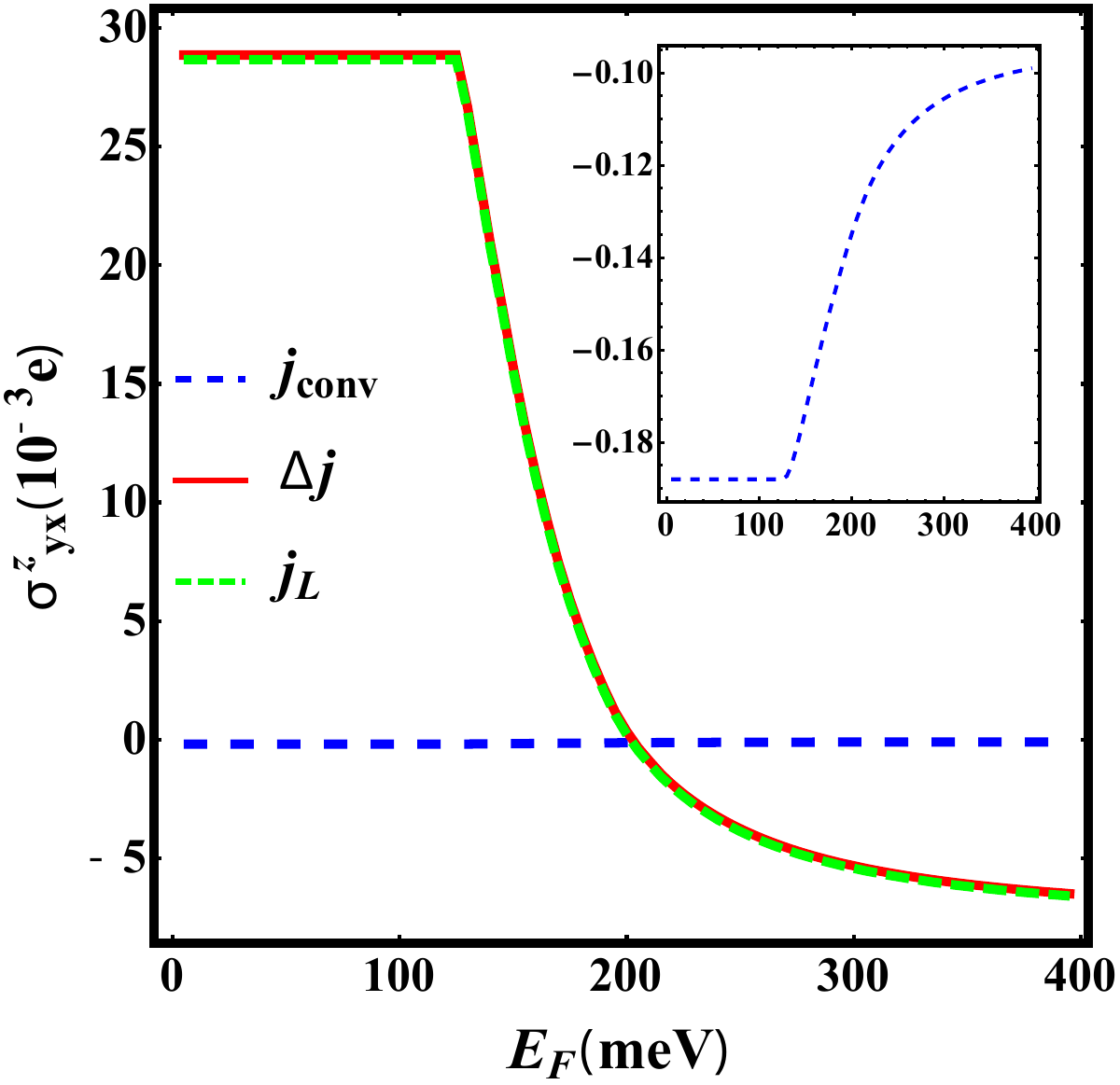}
\caption{\label{CuMnAs-int}
The intrinsic OHE for CuMnAs [Eq.~\ref{model}] with OAM polarization along $\hat{\bm z}$ and electric field along $\hat{\bm x}$. The inset plots the conventional contribution $j_{\rm conv}$ again, with a different scale on the y-axis. The parameters are $t = 0.08\,{\rm eV}$ and $\tilde{t} = 1\,{\rm eV}$, $\alpha_{\rm R}=0.8,\alpha_{\rm D} =0$,and ${\bm h}_{\rm AFM} =(0.85,0,0)\,{\rm eV}$.}
\end{center}
\end{figure}

In this paper we present a full quantum mechanical evaluation of non-equilibrium orbital current accounting for all matrix elements of the position operator, including its intra-band matrix elements, which are technically challenging. Our principal finding is the full expression for the orbital current in response to an electric field, given by $j\nd_L = j_{\rm conv} + \Delta j$. Here $j_{\rm conv}$ is the conventional orbital current, which has been evaluated to date, and is based on the equilibrium OAM and the non-equilibrium distribution. The \textit{quantum corrections} $\Delta j$ consist of contributions from: (i) the electron group velocity leading to band off-diagonal matrix elements of the OAM; (ii) inter-band coherence terms induced by the band-diagonal part of the position operator; and (iii) the non-commutativity of the position and velocity operators. We develop a quantum kinetic equation for the density matrix and determine a general expression for the OHE applicable to an arbitrary band structure, focussing on the intrinsic case. Our result, presented in Eqs.~\ref{Jconc}-\ref{Jcorr} below, is general and applies to all solids, having important implications for the interpretation of experiments. To illustrate this, we first evaluate the full OHE for the topological antiferromagnet CuMnAs, shown in Fig.~\ref{CuMnAs-int}. It is seen that $\Delta j$ provides the dominant contribution to the OHE, and the resulting OHE has the opposite sign to the conventional formula. We also evaluate the full OHE for a model of massive Dirac fermions, where again the quantum corrections dominate the response, and are also influenced by tilting and warping effects. We stress that our concern is with the \textit{evaluation} of the orbital current, not its definition, which we take to be the usual product of the OAM and velocity operators. 

{\color{blue}\textit{OAM and orbital current operators.}} We begin by considering a generalized dipole operator $\HBd^A = \half\{ \hat{A}, \HBr \}$, where $\hat{A}$ represents a general operator. The expectation value of such an operator is Tr$(\HBd^A\hat{\rho})$ with $\hat{\rho}$ the density matrix. We define the OAM operator as the symmetrized combination $\hat{\bm L}=\frac{1}{2}(\hat{\bm r}\times\hat{\bm v} - \hat{\bm v}\times\hat{\bm r})$, with $\hat{\bm v}$ the velocity operator, and we work in the Hilbert space spanned by Bloch wave-functions $\ket{\Psi_{m{\bm k}} } = e^{i{\bm k}\cdot{\bm r}} \ket{u_{m{\bm k}}}$. Determining the OAM expectation value is tantamount to substituting the appropriate velocity component for $\hat{A}$ above, whereupon the correct expression for the equilibrium OAM in the modern theory immediately emerges. 

The orbital current operator is $\Hj^\alpha_\delta = \half\big\{\HL_\alpha,\Hv_\delta\big\}$ and its expectation value is $\Tra\,(\Hj^\alpha_\delta\,\Hrho)$. Using the cyclic property of the trace we decompose the orbital current operator into a dipole-like part and a commutator part as follows:
\begin{equation}\label{operator-OHE}
j^\alpha_\delta=\frac{1}{8}\epsilon_{\alpha\beta\gamma}\Tra\Big(\big\{\{ \Hr_\beta,\hat{\rho}\},\{\Hv_\gamma,\Hv_\delta\}\big\} +  \big\{\Hrho,\big[\Hv_\gamma,[\Hr_\beta,\Hv_\delta]\big]\big\}\Big)\ .
\end{equation}
This decomposition helps fix the conventional orbital current and shed light onto the additional terms. The first term contains the conventional orbital current $j_{\rm conv}$. This can be obtained immediately by retaining only the inter-band terms in $\Hr_\beta$, $\Hv_\gamma$ and $\Hv_\delta$. The \textit{quantum correction} stems from: (i) the band-diagonal terms in the velocity operators; (ii) the band-diagonal terms in the position operator, and (iii) the commutator $[\hat{r}_\beta,\hat{v}_\delta]$ in the second term, which reflects the non-commutativity of the position and velocity operators. In the crystal momentum representation the commutator matrix elements are
\begin{equation}
i\big[\Hr_\beta,\Hv_\delta\big]^{nm}_{\bm k} = \bigg(\der{v_\delta}{k_\beta}\bigg)^{\!\!nm}_{\!\!\Bk} -i\,\big[\CR_\beta, v_\delta\big]^{nm}_{\bm k} \equiv \bigg(\Der{v_\delta}{k_\beta}\bigg)^{\!\!nm}_{\!\!\Bk}\ ,
\end{equation}
where $\CR_\beta^{lm} = i\,\sbraket{u_l}{\der{u_m}{k_\beta}}$ is the Berry connection, which reflects the wave vector dependence of the basis functions. The velocity operator in turn has matrix elements $\hbar\,[\Hv_\delta]^{mn}_{\bm k}=\partial\ve_m/\partial k_\delta\, \delta_{mn} + 
i\,\CR^{mn}_\delta (\ve_m-\ve_n)$, where $\ve_m=\ve_{m\Bk}$ is the band energy at wavevector $\Bk$.


{\color{blue}\textit{Orbital Hall Effect: General expression.}} To evaluate the OHE we require the non-equilibrium correction to the density matrix in an electric field, for which we turn to linear response theory following the approach of Refs.~\cite{Interband-Coherence-PRB-2017-Dimi, JE-PRR-Rhonald-2022}. The single-particle density operator obeys the quantum Liouville equation, $\pz\Hrho/\pz t + (i/\hbar)[\HH,\Hrho]=0$, where $\HH=\HH_0+\HH_E$. Here $\HH_0$ is the band Hamiltonian and $\HH_E=e\BE\cdot\HBr$ is potential due to the external electrical field. At this stage we focus on intrinsic effects and do not consider disorder scattering. In the crystal momentum representation the equilibrium density matrix has the diagonal form $\rho_{0\Bk}^{mn} = f_m\, \delta_{mn}$, where $f_m \equiv f(\ve_{m{\bm k}})$ is the Fermi-Dirac distribution for band $m$. In an electric field the density matrix can be written as $\hat{\rho} = \rho_0+\rho_E$, and, in linear response, it has been shown that \cite{Interband-Coherence-PRB-2017-Dimi} 
\begin{equation}\label{rhoE}
\rho^{mn}_{E{\bm k}} = {f(\ve_{m{\bm k}})-f(\ve_{n{\bm k}})\over
\ve_{m{\bm k}} -\ve_{n{\bm k}} }\> e{\bm E}\cdot \BCR^{mn}_\Bk\quad.
\end{equation}

\begin{figure}[t]
\begin{center}
\includegraphics[trim=0cm 0cm 0cm 0cm, clip, width=0.9\columnwidth]{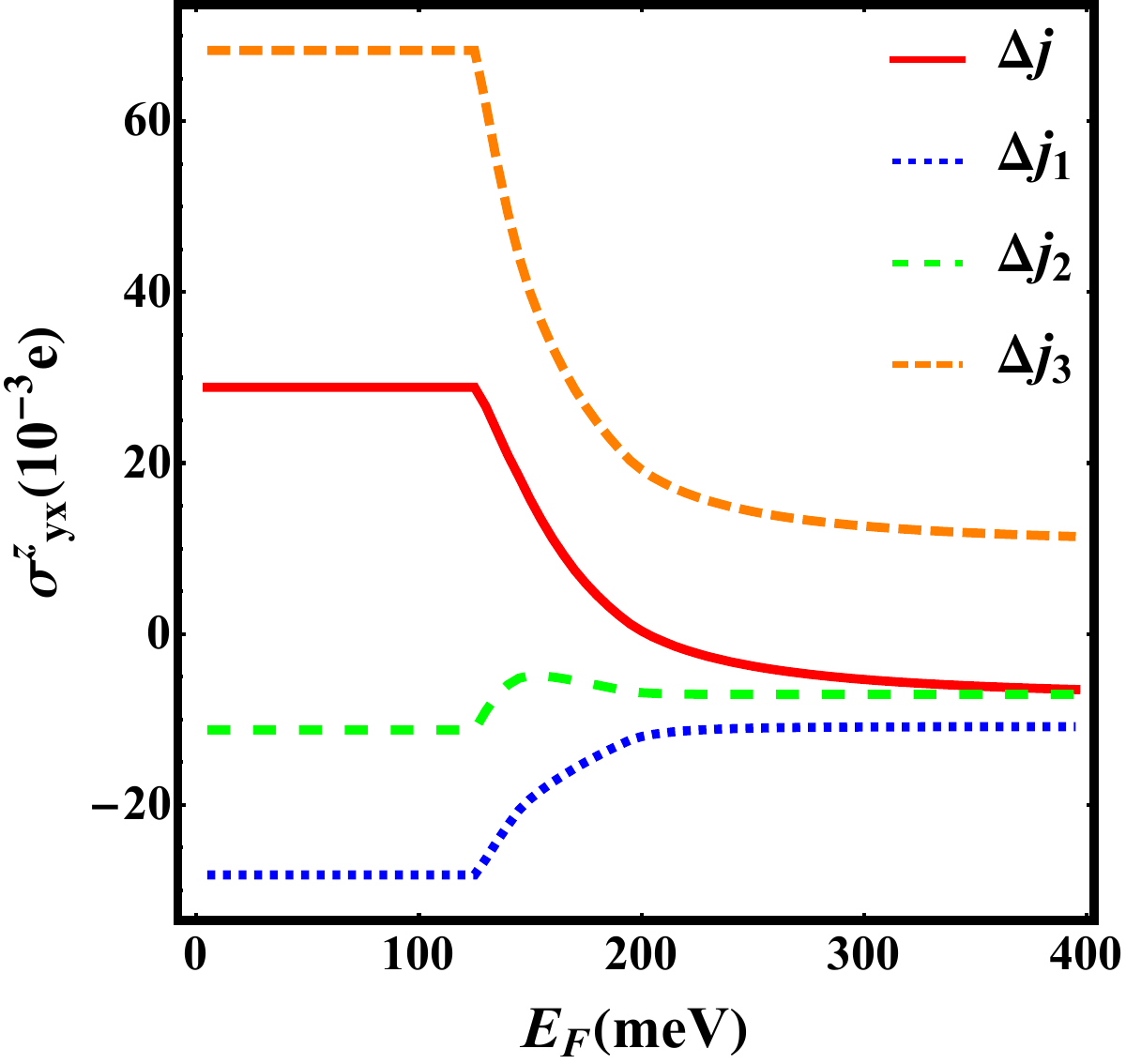}
\caption{\label{CuMnAs-qm}
The decomposed quantum correction $\Delta j$ for CuMnAs with OAM polarization along $\hat{\bm z}$ and electric field along $\hat{\bm x}$. We have used the same parameters as in Fig.~\ref{CuMnAs-int}. }
\end{center}
\end{figure}

Once $\rho^{mn}_{E{\bm k}}$ is found the conventional orbital current is
\begin{equation}\label{Jconc}
    j_{\rm conv} \displaystyle = \frac{1}{4}\eps_{\alpha\beta\gamma}\sum_{m,\Bk} \big\{\CR\nd_\beta, \rho\nd_{E\Bk}\big\}^{mm}\, \big\{v_\delta, v_\gamma\big\}^{mm}\,,
\end{equation}
where only band off-diagonal components of the velocity operators enter. The \textit{quantum corrections} can be written as $\Delta j = \Delta j_1 + \Delta j_2 +\Delta j_3$, where
\begin{align}\label{Jcorr}
\Delta j^\alpha_{\delta,1} &= \half\eps_{\alpha\beta\gamma} \sum_{m,\Bk} \big\{\CR\nd_\beta, \rho\nd_{E\Bk}\big\}^{mm}\, v^{mm}_\delta v^{mm}_\gamma \\
\Delta j^\alpha_{\delta,2} &=\frac{i}{4}\eps_{\alpha\beta\gamma}\!\!\!
\sum_{m,n,\Bk} \!\!\! {2eE_\mu \left[\Der{\Xi^0_\beta}{k_\mu}\right]^{mn}\!\!\!\!
+\big\{\hbar v\nd_\beta, \rho\nd_{E\Bk}\big\}^{mn} \over \ve_n-\ve_m}
\{v_\gamma, v_\delta\}^{nm}\nonumber\\
\Delta j^\alpha_{\delta,3} &= \frac{i}{4}\eps_{\alpha\beta\gamma}\, \!\!\!
\sum_{m,n,\Bk} \Big[v_\gamma, \Der{v_\delta}{k_\beta} \Big]^{mn}_\Bk \rho^{nm}_{E\Bk} \quad,\nonumber
\end{align}
where we have abbreviated $\big[\Xi^0_\beta\big]^{mn}=\half\CR^{mn}_\beta (f_m + f_n)$, and $m \ne n$ is understood in all the summations. These expressions are \textit{general} and apply to arbitrary band structures, provided the velocity and Berry connection matrix elements can be calculated. The result is gauge invariant, as shown in the Supplement\cite{Supplement}. The OAM polarization is taken to be along the $\alpha$-direction while the transport direction is denoted by $\delta$. We stress that $\Delta j$ has never been reported previously. The three terms entering the quantum correction $\Delta j$ reflect inter-band coherence induced by the electric field. $\Delta j_1$ has the same structure as $j_{\rm conv}$, except that the electron group velocity appears instead of the inter-band velocity. $\Delta j_2$ is the result of restoring the band-diagonal matrix elements of the position operator on the left side of Eq.~\ref{operator-OHE}. Finally, $\Delta j_3$ reflects the non-commutativity of the position and velocity operators.

{\color{blue}\textit{Tetragonal CuMnAs.}} We now exemplify our evaluation by studying the OHE in a tight-binding model of CuMnAs. CuMnAs can exist in both orthorhombic and tetragonal crystal structures \cite{Ter-CuMnAs,Tet-CuMnAs-Sci-Rep}, both can host the antiferromagnetic Dirac semimetal phase. Here we take the stable tetragonal CuMnAs phase as an example, which has opposite spins lying on a bipartite lattice. Such an arrangement preserves the combined $\mathcal{PT}$ symmetry by the exchange of the sublattices with the flip of oppositely aligned spins \cite{PhysRevB.108.L201405, CuMnAs-PhysRevX.11.011001}, enforcing that bands at each momentum are doubly degenerate \cite{Review-Topolocial-Semimetal}. The tight-binding Hamiltonian for CuMnAs without external magnetization is written as \cite{XiaoDi-PhysRevLett.127.277201}
\begin{equation}\label{model}
H_{0,\Bk}=\begin{pmatrix}
\ve_0(\Bk) + \Bh(\Bk)\cdot\Bsigma  & V_{\rm AB}(\Bk) \\
V_{\rm AB}(\Bk) & \ve_0(\Bk)- \Bh(\Bk)\cdot\Bsigma
\end{pmatrix}\quad,
\end{equation}
where $\ve_0({\bm k})=-t(\cos k_x+\cos k_y)$ and $V_\text{AB}=-2\tilde{t}\cos(k_x/2)\cos(k_y/2)$, where $t$ and $\tilde{t}$ denote hopping between orbitals of the same and different sublattices, respectively. The sublattice-dependent spin-orbit coupling and the magnetization field are included in 
\begin{equation}
\begin{split}
\Bh(\Bk)&=\big(h^x_\AFM -\alpha\nd_\RR\sin k_y+\alpha\nd_\RD 
\sin k_y \,, \\
&\hskip0.5in h^y_\AFM +\alpha\nd_\RR\sin k_x+\alpha\nd_\RD \sin k_x 
\, , \,h^z_\AFM\big)
\end{split}
\end{equation}
with $\alpha\nd_\RR$ and $\alpha\nd_\RD$ the Rashba and Dresselhaus type spin-orbit coupling coefficients \cite{CuMnAs-TB-PhysRevX.11.011001}. In the band dispersion gapless points appear along the high symmetry line $(k_x=\pi)$. We expand the Hamiltonian in the vicinity of $(k_x,k_y ) = (1, 0.5)\pi$ up to $\delta k_{x,y}=\pm0.1\pi$. Then we numerically integrate the full OHE for the approximated model Hamiltonian around $(1, 0.5)\pi$ up to $\pm0.1\pi$ by ${\bm k}$-discretization. This is justified since the OAM itself is strongly peaked around these points. We calculate the orbital Hall conductivity dependence on the Fermi energy $E_\RF$, when $E_\RF$ is in the conduction band, $E_\RF=0$ corresponds to the middle of the band gap. For CuMnAs we obtain the total intrinsic OHE in Fig.~\ref{CuMnAs-int}. \footnote{Note: our plotted values of the orbital conductivity have been multiplied by $-1$, this is for the added factor of the electron charge in the orbital magnetic moment, such that our results are comparable with Ref.~\cite{IOHE-PRB-2021-Giovanni}.}. Remarkably, the conventional contribution corresponds to only one fraction of the total orbital current, while the quantum correction $\Delta j$ accounts for almost the entire value, which has the opposite sign. To determine the sign the OHE could in principle be imaged using the techniques developed for OAM densities \cite{Exp-BC-NatPhy-2022-Eli}. The current is non-zero in the gap due to the Fermi sea contribution from the filled valence band states. We also plot the individual contributions to $\Delta j$ in Fig.~\ref{CuMnAs-qm}, showing that in this model the dominant term is $\Delta j_3$, which arises from the non-commutativity of the position and velocity operators.

\begin{figure}[tbp]
\begin{center}
\includegraphics[trim=0cm 0cm 0cm 0cm, clip, width=0.9\columnwidth]{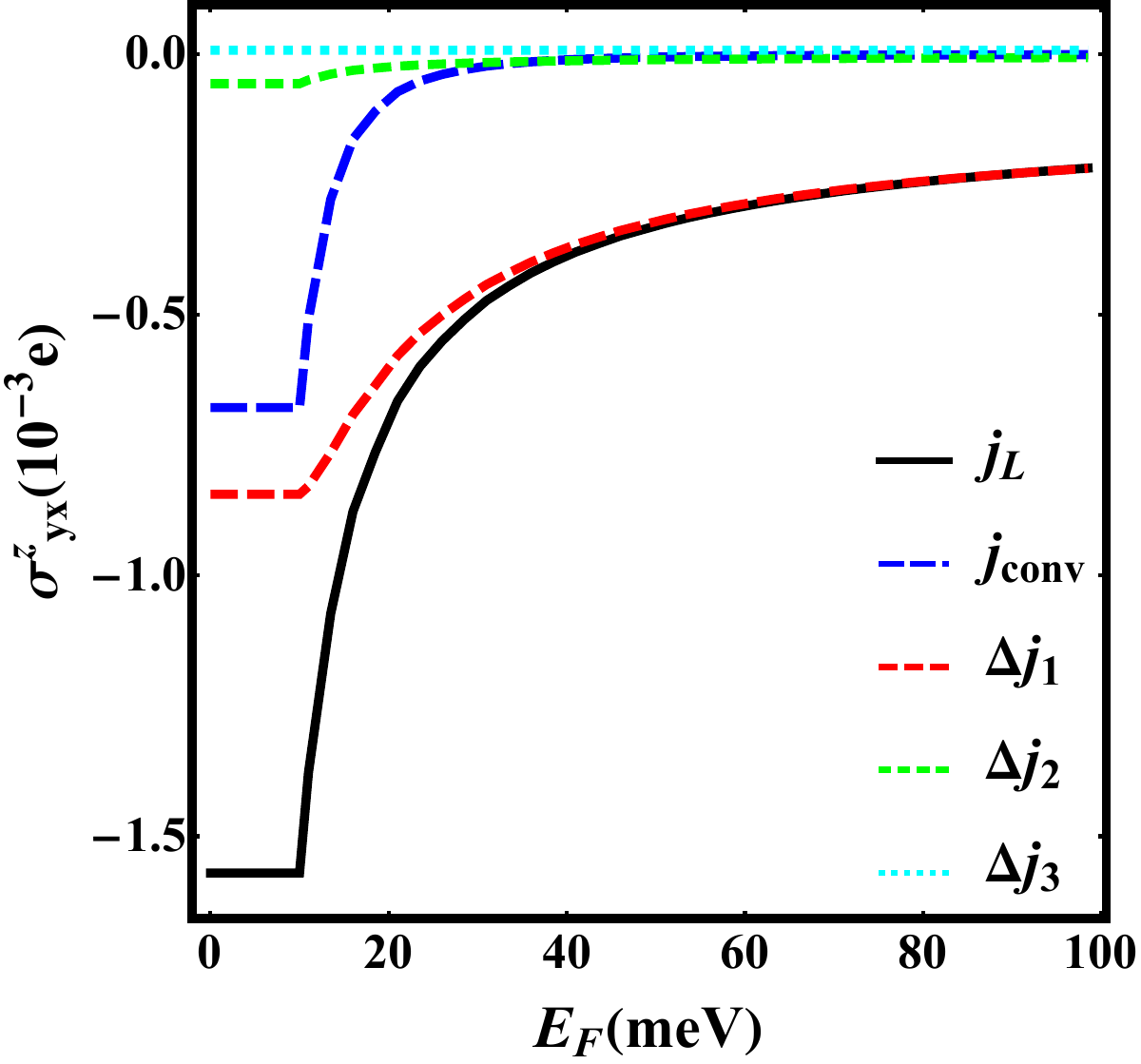}
\caption{\label{Dirac_plot}
The intrinsic $j_L$ and each of its components for our two-band Dirac model. Here we use $\alpha = 4$ eV{\AA}, $m = 0.01$ eV, $t = 0.2 \alpha$ and $\kappa = 0.8$ eV{\AA}$^2$.}
\end{center}
\end{figure}

{\color{blue}\textit{Dirac Model.}} We look to further emphasise the importance of the quantum correction in the orbital current by examining a simple and general two-band model -- a massive tilted Dirac cone with trigonal warping
\begin{equation}
\begin{aligned}
    H_{0,\bm k} =& \alpha(k_y \sigma_x - k_x \sigma_y) + \kappa((k_x^2 - k_y^2)\sigma_x - 2k_x k_y \sigma_y) \\
    &+ m \sigma_z + t k_x \sigma_0\,,
\end{aligned}
\end{equation}
where $\kappa$ is the warping parameter, $t$ is the tilting parameter and, $\sigma_{x,y,z,0}$ are the Pauli matrices. If one first ignores the warping and tilt terms, then for the Fermi energy $E_\RF$ in the conduction band the orbital current is
\begin{equation}
\begin{split}
j_{{\rm conv},y}^z &= {e E_x m^2 \alpha^2\over 24 \pi \hbar^2 E_\RF^3}\\
\Delta j_{1,y}^z &= {e E_x \alpha^2(3E_\RF^2 - m^2)\over 48 \pi \hbar^2 E_\RF^3}\quad.
\end{split}
\end{equation}
In this simple case $\Delta j_{2,y}^z$ and $\Delta j_{3,y}^z$ are zero. At the conduction band bottom $E_F\rightarrow m$ and the two contributions are equal. However, as the Fermi energy increases the conventional contribution decays rapidly, whereas the quantum correction decays much more slowly and dominates. When $E_\RF$ is in the gap the orbital current will be quantized, the total orbital current will be $j_L = e E_x \alpha^2/ 12 \pi \hbar^2 m$. The same behaviors are also seen in Fig.\ref{Dirac_plot} when including both the warping and tilt. When warping and the tilt are included we find $\Delta j_{2,y}^z$ and $\Delta j_{3,y}^z$ to be non-zero, yet they remain relatively small. Here $\Delta j_{1,y}^z$ is the dominant contribution to the orbital current. Hence, even for this simple model we find that the conventional method for the evaluation of the orbital current is insufficient and the quantum correction is required for an accurate calculation.

{\color{blue}\textit{Discussion.}} We have shown that the non-equilibrium expectation value of the orbital current operator contains terms that have not been evaluated previously, which, for the models considered, have substantial magnitudes. How does our approach differ from what has been done to date? To answer this, we note that the starting point of any quantum mechanical calculation is the operator under consideration. The OAM itself is the product of the position and velocity operators. The equilibrium OAM stems from the inter-band matrix elements of these operators, which were traditionally neglected \cite{Ashcroft76, cardona1969solid}, and the OAM would have continued to be missed if this neglect persisted. According to the same reasoning, the starting point of any calculation of the OHE must be the orbital current \textit{operator}, that is, the product of the OAM and velocity operators. The non-equilibrium expectation value of the orbital current operator will contain intra-band and inter-band matrix elements of both the position and velocity operators. To date, only the inter-band position matrix elements were retained in the OHE, while the intra-band position matrix elements were neglected. 

The reason for this neglect is straightforward. Quantum mechanically, in principle, once the matrix elements are known the operator is known. It is natural to assume that constructing the orbital current operator matrix elements involves multiplying the known OAM matrix elements with the known velocity matrix elements, and extracting the OAM matrix elements from the equilibrium expectation value of the OAM. This procedure, however, is problematic, because of two reasons. Firstly, the matrix elements of the position operator for Bloch electrons are differential operators. Secondly, the equilibrium density matrix only selects one part of the OAM matrix elements. The equilibrium OAM expectation value involves only off-diagonal matrix elements of the position and velocity operators \cite{GaneshSundaram1999, Mingche-1995-PhysRevLett.75.1348, Mingche-1996-PhysRevB.53.7010}. The net result is that intra-band matrix elements of the position operator are thrown away when the OAM expectation value is taken in equilibrium. Hence, the OAM matrix elements extracted from the equilibrium OAM expectation value are not sufficient to determine the orbital current. The intra-band matrix elements of the position operator must be included, and we have developed a powerful technique for handling them. We have focussed on the Bloch picture of electrons in crystals. Different insights into the OAM emerge from the Wannier and Bloch formulations \cite{Rhonald-Rev, BiTMD-OHE-PRB-2022-Giovanni&Tatiana, OHE-PRB-2022-Manchon, PhysRevB.108.075427-Manchon, PhysRevResearch.5.043052, Quek-PhysRevResearch.2.033256}. Since an exact correspondence exists between the Bloch and Wannier bases, the fundamental considerations outlined in this work will be applicable to Wannier studies as well. 

What is the physical meaning of the quantum corrections to the OHE? We recall that the discussion of Bloch electrons' position is entirely based on an equilibrium picture, and the equilibrium location of an electron in the lattice is undefined. Nevertheless, out of equilibrium an electric field in general displaces an electron's center of mass away from its equilibrium location, generating a dipole. In the intrinsic case, electric-field induced changes to the density matrix are purely band off-diagonal, this dipole represents a DC version of the inter-band polarization known in non-linear optics \cite{Sipe-Non-linear,Vanderbilt-polarization-PhysRevB}. This suggests one mechanism behind the OHE as the generation of an inter-band polarization by the applied electric field: the electric field generates a dipole which then rotates about the centre of mass, generating an OAM, which is convected along with the electron. This is the physics behind $\Delta j_1$ \footnote{In incorporating the band-diagonal position matrix elements one must ensure that no part of the final result depends on the equilibrium centre of mass of the electron. Our boundary conditions take this into account.}. Next, the OAM operator itself has inter-band matrix elements. These do not contribute to the equilibrium OAM expectation value, because the equilibrium density matrix is diagonal in the band index. Nevertheless, the inter-band matrix elements of the OAM operator do contribute to OAM transport -- they contribute to the OHE. It is these matrix elements that are primarily responsible for $\Delta j_2$. Finally, once the position and velocity are treated as full operators, one must also account for the fact that they do not commute, and this leads directly to $\Delta j_3$. In equilibrium, this commutator can be related to the effective mass,\cite{sun2024theory} yet out of equilibrium the resulting expression is considerably more complex. Disentangling the different contributions experimentally is challenging. However, in the case of CuMnAs the predictions differ by three orders of magnitude, so a comparison with experimental results, once they become available, can be made unambiguously.

We stress that our work is concerned with the evaluation of the current, not its definition. In the long run the same fundamental issues will need to be considered for the OHE as for the spin-Hall effect. One is that the orbital Hall current, as defined, can be nonzero in equilibrium in the same way as the spin current \cite{Rashba-PhysRevB.68.241315}. In fact all the four contributions to $j_L$ could be separately nonzero in equilibrium, although they do vanish for all the effective models we have investigated -- CuMnAs, bulk topological insulators, and massive Dirac fermions. Secondly, the OAM itself may not be conserved in an electric field, as we recently showed \cite{Rhonald-Conservation-OMM}, and one may eventually need to consider a \textit{proper} OH current in analogy with the proper spin current. Thirdly, the only observable is the OAM, and the relationship between the current and the OAM accumulation at the boundary will have to be determined. Nevertheless, whichever approach is taken, the fundamental problem of the evaluation of $j_L$ is relevant. 

{\color{blue}\textit{Conclusion.}} We have presented a full quantum mechanical evaluation of the OHE, demonstrating that the existence of quantum corrections that have been missed in conventional evaluations. These corrections stem from the band-diagonal terms in the position and velocity operators, as well as the non-commutativity of the position and velocity operators. In the models considered the quantum corrections are at least as large as the conventional terms in the OHE, and in CuMnAs they overwhelm the conventional result. 

\textit{Acknowledgments} -- This work is supported by the Australian Research Council Centre of Excellence in Future Low-Energy Electronics Technologies, project number CE170100039. JHC acknowledges support from an Australian Government Research Training Program (RTP) Scholarship. We acknowledge enlightening discussions with Rhonald Burgos, Qian Niu, Naoto Nagaosa, Guang-Yu Guo, Hsin Lin, Mikhail Titov, Aires Ferreira, Tatiana Rappoport, and Giovanni Vignale.

\end{document}